\documentclass[journal=ancac3,manuscript=article]{achemso}

\usepackage[version=3]{mhchem} 

\title{Site-selective ion beam synthesis and optical properties of individual \ce{CdSe} nanocrystal quantum dots in a \ce{SiO2} matrix}

\author{H. Moritz Mangold}
\affiliation[EP1, Univ. Augsburg]{Emmy Noether Junior Reserach Group at Lehrstuhl f\"{u}r Experimentalphysik 1 and Augsburg Centre for Innovative Technologies (ACIT), Universit\"{a}t Augsburg, Universit\"{a}tsstr. 1, 86159 Augsburg, Germany}
\author{Helmut Karl}
\affiliation[EP IV, Univ. Augsburg]{Lehrstuhl f\"{u}r Experimentalphysik IV, Universit\"{a}t Augsburg, Universit\"{a}tsstr. 1, 86159 Augsburg, Germany}
\author{Hubert J. Krenner}\email{hubert.krenner@physik.uni-augsburg.de}
\affiliation[EP1, Univ. Augsburg]{Emmy Noether Junior Reserach Group at Lehrstuhl f\"{u}r Experimentalphysik 1 and Augsburg Centre for Innovative Technologies (ACIT), Universit\"{a}t Augsburg, Universit\"{a}tsstr. 1, 86159 Augsburg, Germany}
\alsoaffiliation[CeNS]{Center for NanoScience ({\it CeNS}), Ludwig-Maximilians-Universit\"at M\"unchen, Geschwister-Scholl-Platz 1, 80539 M\"{u}nchen, Germany}


\begin{document}

\begin{abstract}
Cadmium selenide nanocrystal quantum dots (NC-QDs) are site-selectively synthesized by sequential ion beam implantation of selenium and cadmium ions in a \ce{SiO2} matrix through sub-micron apertures followed by a rapid thermal annealing step. The size, areal density and optical emission energy of the NC-QDs are controlled by the ion fluence during implantation and the diameter of implantation aperture. For low fluences and small apertures the emission of these optically active emitters is blue-shifted compared to that of the bulk material by $>100\,{\rm meV}$ due to quantum confinement. The emission exhibits spectral diffusion and blinking on a second timescales as established also for solution synthesized NC-QDs.
\end{abstract}

Over the past three decades the investigation of nanocrystal quantum dots (NC-QDs)\cite{Brus:84} has been driven by the broad variety of different practical applications and gaining fundamental insights on the physical principles giving rise to their unique properties. Since the monodisperse NC-QDs can be synthesized in solutions \cite{Murray1993} and intentionally doped with impurities, novel opto-electronic \cite{Norris2008}, biomedical \cite{Klostranec2006} and photovoltaic applications \cite{Kramer2011} came within reach. While the straightforward solution synthesis represents a major advantage for these applications, NC-QD composites with other materials have to be realized in a top-down solution based approach. A post-fabrication hybridization of a host material by embedding NC-QDs is extremely challenging. Such hybrid systems would be particularly suited for NC-QD single photon sources \cite{Michler2000} for quantum cryptography applications or NC-QD based spintronics\cite{Berezovsky:06a}.
One alternative approach to synthesize nanocrystals (NCs) and NC-QDs in almost any solid state substrate is based on a standard technique in semiconductor industry, ion beam implantation. Here the elements of a desired compound are implanted in the correct stoichiometry and the NCs are formed during a subsequent temperature treatment. By this technique, a wide variety of elemental and compound NCs can be synthesized in planar and non-planar substrates\cite{Meldrum2001,Karl2005}. In addition to optically passive NCs \cite{Lopez:02a,Zimmer2012}, ensembles of optically active NC-QDs of elemental semiconductors, most notably silicon \cite{Min1996}, and II-VI compounds \cite{Hipp2003} have been synthesized and used as active materials in optoelectronic and photonic devices\cite{Pavesi2000,Walters2005,Achtstein2006,Kippenberg2009}. Since ion beam implantation is typically performed on large areas, synthesis and spectroscopy of isolated NC-QDs has turned out to be extremely challenging \cite{Valenta2004}.\\
In this letter we report on the synthesis of individual, optically active cadmium selenide NC-QDs by sequential ion beam implantation through sub-micron implantation masks. We show that the combination of low ion fluences and feature sizes smaller than 1000\,nm can be readily used to synthesize small diameter NC-QDs with quantum confined electronic states at low areal density. These NC-QDs exhibit sharp spectral lines which can be tuned spectrally during fabrication by the implantation aperture diameter. Single NC-QD emission lines exhibit both spectral diffusion and blinking (intermittency) which are characteristic fingerprints also observed for their colloidal counterparts.  \\

\section{Sample fabrication and SEM characterization}

\begin{figure}[htb]
    \begin{center}
       \includegraphics[width=0.85\columnwidth]{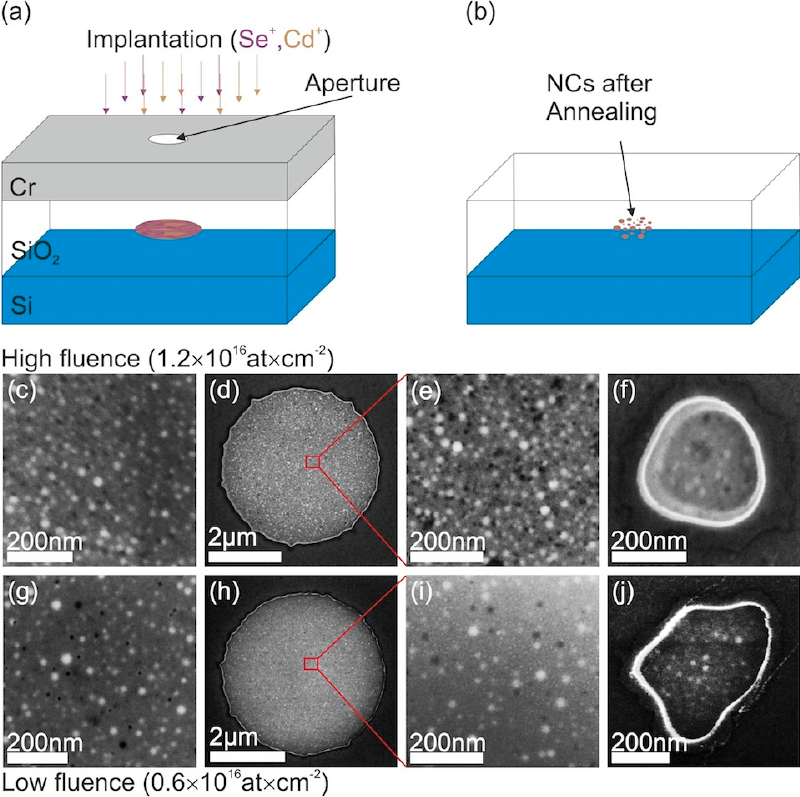}
        \caption{Synthesis and SEM characterization -- (a+b) Schematic of site-selective ion beam synthesis through sub-micron apertures. SEM images of high (upper row) and low fluence (bottom row) samples from unpatterned regions (c,g), large apertures (d+e,h+i) and small apertures (f,j)}
        \label{fig:1}
    \end{center}
\end{figure}

The experimental procedure to synthesize \ce{CdSe} NCs is presented in Fig. \ref{fig:1} (a) and (b) and consists of two major steps. First implantation apertures are defined by electron beam lithography and wet chemical etching in a 120\,nm thick, evaporated chromium mask. The diameters of these apertures $(d)$ ranged between $d=400\,{\rm nm}$ and $d=4500\,{\rm nm}$. An unpatterned region without a \ce{Cr} mask provides a reference on the same substrate. In the following, \ce{Se+} followed by \ce{Cd+} ions were implanted at cryogenic temperature (77\, K) in the 200\, nm thick \ce{SiO2} thermal oxide on top of a commercial \ce{Si} wafer with energies of 134\,kV, and 190\,kV, respectively as shown in Fig. \ref{fig:1} (a). These energies give rise to Gaussian concentration profiles for both ion species centered at a projected range of 100\,nm below the sample surface. Here we present results from two different substrates implanted with ion fluences of $F_0= 0.6\cdot 10^{16}\,\text{at}/\mathrm{cm^{{2}}}$ and $2F_0$, which we refer to as the low and high fluence sample in the following. We want to note that these fluences are significantly reduced compared to previous work in which no indications of a quantum confined level structure has been observed due to a large NC size \cite{Achtstein2006}. In the second and final step depicted in Fig. \ref{fig:1} (b), the \ce{Cr} implanation mask is removed by wet chemical etching and NCs are formed in the previously unmasked region during a rapid thermal annealing step at $900^\circ\text{C}$ for 30\,s.\\
 
We characterized the such synthesized NCs samples by scanning electron microscopy. Typical micrographs from an unpatterned region are compared to that from large and small apertures in \ref{fig:1} (c-f, high fluence sample) and (g-j, low fluence sample). In images (d),(f),(h) and (j) the implanted areas are clearly resolved as bright regions due to the improved contrast of the \ce{CdSe-SiO2} composite. Individual NCs (bright) and voids (black) are clearly visible at high magnification in unpatterned regions in (c) and (g), inside large $d\sim 4000\,{\rm nm}$ apertures (e) and (i) and in the micrographs recorded sub-micron apertures (g) and (j). No NCs are visible in the masked areas outside the apertures proving the site-selectivity of our approach. The high magnification images from unpatterned regions (c) and (g) clearly demonstrate that the areal density and size of both NCs (bright) and voids (black) are clearly reduced for the low fluence sample when compared to the high fluence sample. This is expected due to the reduced amount of \ce{Cd} and \ce{Se} introduced. While high resolution SE micrographs from unpatterned regions and inside large apertures show similar NC sizes and areal densities for the high fluence sample [cf. \ref{fig:1} (c) and (e)], the same analysis yields a reduction of both quantities when comparing Fig. \ref{fig:1} (g) and (i) from the low fluence sample. This finding provides a first indication that implantation even through micron-sized apertures gives rise to a modified NC synthesis. When decreasing the aperture size further, the micrographs in Fig. \ref{fig:1} (f) and (j) show, that for both fluences, the average NC size decreases. In particular, for the low fluence sample, NC diameter range at or below the spatial resolution. For such small diameter NCs we expect QD properties.

\section{Ensemble characterization}
\begin{figure}[htb]
    \begin{center}
        \includegraphics[width=0.7\columnwidth]{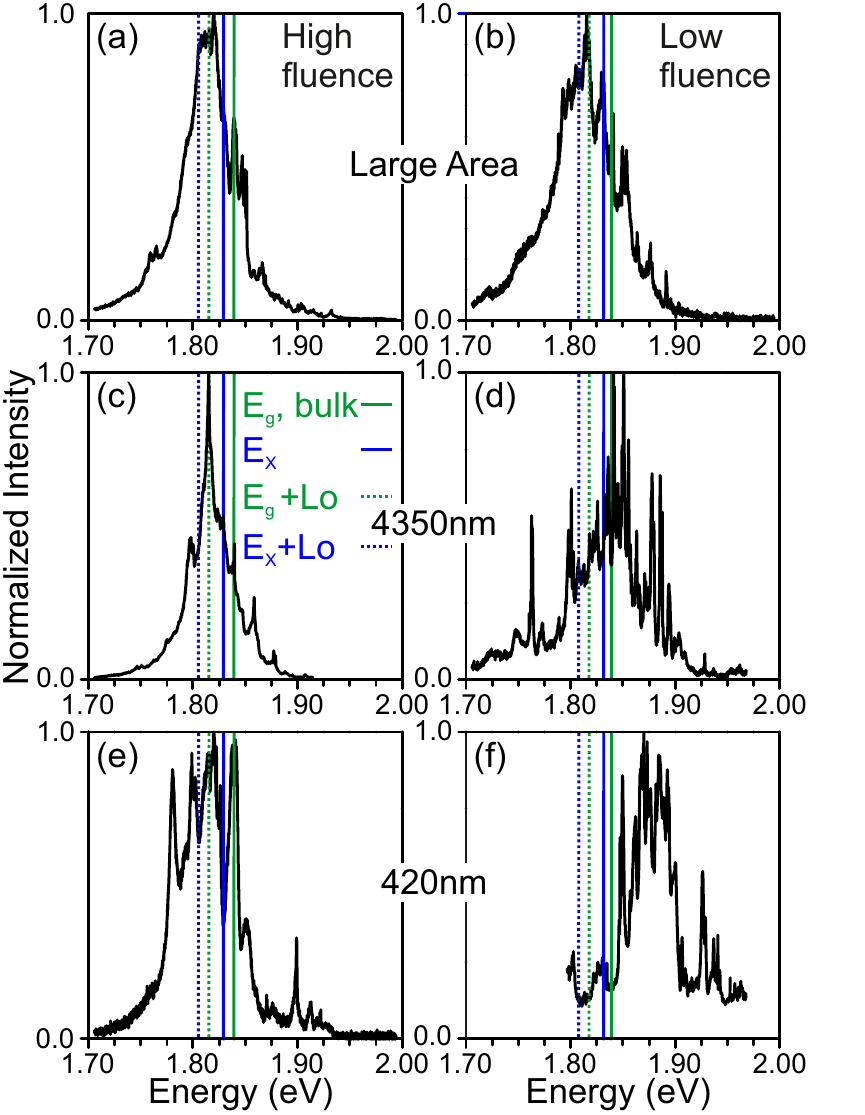}
        \caption{Ensemble optical properties -- Normalized $\mu$-PL spectra from the high (left) and low (right) fluence samples recorded from unpatterned regions (a,b), large apertures (c,d) and small apertures (e,f). To assess ensemble averages (e) and (f) show integrations of spectra recorded from 5 and 10 implantation sites, respectively. The bulk \ce{CdSe} at $T=8.5\,{\rm K}$ is marked by green lines. The red (blue) lines mark the center of the ensemble emission in the unpatterned region (for $d=420\,{\rm nm}$ apertures).}
        \label{fig:2}
    \end{center}
\end{figure}
\begin{figure}[htb]
    \begin{center}
       \includegraphics[width=0.9\columnwidth]{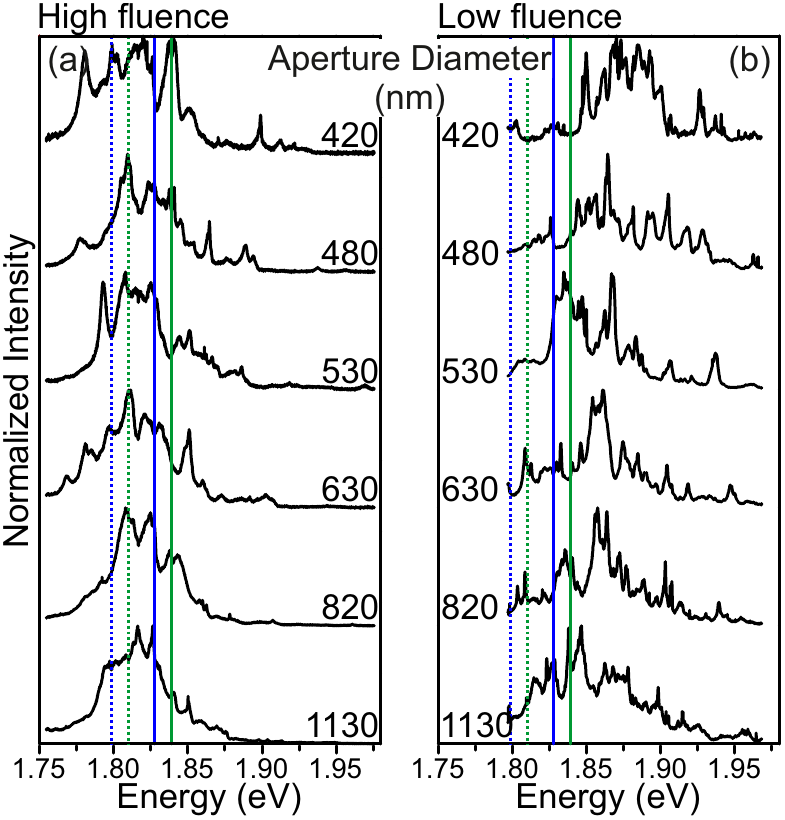}
        \caption{Ensemble optical properties for sub-micron apertures -- Normalized, ensemble averaged $\mu$-PL spectra from the high (left) and low (right) fluence samples recorded from aperture diameters ranging from $d=430$-$1130\,{\rm nm}$. Data is integrated from 5 and 10 implantation sites on the high and low fluence sample, respectively. The bulk \ce{CdSe} at $T=8.5\,{\rm K}$ is marked by green lines. The red (blue) lines mark the center of the ensemble emission in the unpatterned region (for $d=420\,{\rm nm}$ apertures).}
        \label{fig:3}
    \end{center}
\end{figure}

We expect that the optical properties of our ion beam synthesized NCs reflect their structural size as established for their chemically synthesized QD counterparts\cite{Murray1993,Norris1996,Efros:98}. Therefore, we studied the optical emission of our NCs by conventional low-temperature micro-photoluminescence (PL). The sample was mounted on the coldfinger of a \ce{He}-flow cryostat $(T=8.5\,{\rm K})$ and carriers were photogenerated by a continuous wave diode laser emitting at $\lambda_{\rm Laser}=405 \,{\rm nm}$ focused to a $\sim 1\,\mu{\rm m}$ diameter spot. The NC emission was collected and dispersed by a 0.5\,m single grating monochromator and detected by a cooled \ce{Si} CCD detector. We start by evaluating the ensemble optical properties of these ion beam synthesized NCs. In Fig. \ref{fig:2} we compare the PL of the low and high fluence sample from unpatterend regions on the same sample (a,b), large, $d=4350\,{\rm nm}$ (c,d) and small diameter, $d=420\,{\rm nm}$, (e,f) apertures. To assess the ensemble optical properties of NC fabricated by site-selective implantation through apertures (c-f), we integrated spectra from 5 and 10 different small apertures of identical diameter for the high and low fluence samples, respectively. Furthermore, all PL spectra are normalized to the maximum intensity of the NC emission and the band gap of bulk CdSe of $E_{gap,bulk}=1.839\,{\rm eV}$ at $(T=8.5\,{\rm K})$ is marked by vertical green lines. The spectra recorded from unpatterned regions shown in Fig. \ref{fig:2} (a) and (b) show broad emission peaks which are shifted to lower energies by 23\,meV and 25\,meV compared to $E_{gap,bulk}$ and an inhomogeneous broadening of 43\,meV and 49\,meV for the high and low fluence sample, respectively. The red shift of the NC emission with respect to bulk \ce{CdSe} arises from defects at the \ce{CdSe-SiO2}-interface between the high surface/volume ratio NCs and the surrounding matrix. Furthermore, the observed inhomogeneous broadening can be readily attributed to the size distribution of NCs. For NCs synthesized by site-selective implantation through apertures we observe a dramatically different behavior for the high and low fluence sample. The emission of the high fluence sample remains shifted to lower energies compared to $E_{gap,bulk}$ for both aperture sizes with an increase of the inhomogeneous broadening towards \emph{higher} energies for $d=420\,{\rm nm}$, Fig. \ref{fig:2} (e) compared to $d=4350\,{\rm nm}$, Fig. \ref{fig:2} (c). For all three sets of data of the high fluence sample only weak PL is detected at the high energy side of these spectra, $E>E_{gap,bulk}$, where we expect transitions from NCs exhibiting quantum confined, QD-like spectra (NC-QDs). A close examination of the spectra indeed reveals weak signatures of sharp PL lines in the unpatterned region, Fig. \ref{fig:2} (a) and $d=4350\,{\rm nm}$, Fig. \ref{fig:2} (c). For the smallest implantation aperture size, $d=420\,{\rm nm}$, Fig. \ref{fig:2} (e), the sum spectrum starts to break up in individual sharp lines at $E\sim 1.9\,{\rm eV}$. In contrast, the NC emission of the low fluence sample gets strongly modified when ion implantation is performed through an aperture. For the largest apertures studied, $d=4350\,{\rm nm}$, Fig. \ref{fig:2} (d), the emission is centered close to the bulk band gap of \ce{CdSe} at $E=1.839\,{\rm eV}\sim E_{gap,bulk}$. Most remarkably, it is shifted to the spectral region of quantum confined NCs, $E=1.879\,{\rm eV}$, for the smallest apertures of $d=420\,{\rm nm}$ in Fig. \ref{fig:2} (f). Moreover, in both data of low fluence NCs synthesized through apertures sharp, discrete emission lines are resolvable despite the fact that implantation was performed through a large aperture and spectra from 10 different small apertures were summed. Based on these observations we conclude that (i) a reduction  of the ion fluence favors the formation of smaller NCs exhibiting a quantum confined level structure and (ii) the fraction of NC-QDs can be controllably enhanced by performing the implantation through sub-micron apertures.\\
These conclusions are confirmed further by a detailed investigation of the impact of the aperture diameter $(d)$ on the NC emission properties. In Fig. \ref{fig:3} we present a detailed series of summed NC spectra of the high (a) and low (b) fluence sample for different aperture diameters increasing from $d=1130\,{\rm nm}$ to $d=420\,{\rm nm}$ (same data as in Fig. \ref{fig:2} (e+f)) from the bottom to top spectra. Again spectra of 5 (high fluence sample) and 10 (low fluence sample) different apertures of identical diameter are integrated to assess the ensemble properties. In these plots the center energy of the large area implanted NC spectra is marked by vertical red lines. Comparing the evolution of the NC emission with decreasing $d$ two pronounced differences become apparent between the high and low fluence sample. While for the high fluence sample the center of the emission peak for $d=1130\,{\rm nm}$ agrees with that of the large area implant and only a weak tail towards higher energies is resolved, the NC emission of the low fluence sample is already significantly shifted and broadened towards higher energy. As $d$ decreases the high fluence sample exhibits almost no spectral shift of the center energy and broadens only weakly towards higher energies. In strong contrast, for the low fluence sample both the center energy and the high energy part of the spectrum arising from emission of quantum confined states are shifting systematically towards higher energies.\\

These observations again clearly demonstrate and confirm direct control and correlation of the NCs' optical properties by variation of the total ion fluence and application of a sub-micron implantation apertures. A reduction of the total fluence i.e. the total number of \ce{Se+} and \ce{Cd+} ions introduced per unit area gives rise to the formation of NCs of lower mean diameter. For the low fluence sample in this study we clearly reach the regime of quantum confined NC with QD properties. Moreover, the fraction of these NC-QDs can be enhanced by performing the implantation through a sub-micron aperture. This approach defines the implantation area where ions are incorporated in the \ce{SiO2} matrix. Our experimental finding clearly indicates that this spatial confinement of the implantation area favors the formation of small sized NC-QDs. For the largest diameter apertures $d=4350\,{\rm nm}$ we observe for the low fluence sample a clear shift of the center energy compared to reference NCs in an unpatterned region on the same sample piece. Moreover, for the smallest aperture sizes used in our experiments, $d=420\,{\rm nm}$, the center of the ensemble PL is fully shifted above $E_{gap,bulk}$ indicating that only NC-QDs are formed. Since all diameters are significantly larger than typical ion diffusion lengths during the RTA step in the order of 10's of \,nm, this observation suggests that for low fluences $F\leq F_0$ the introduction of a boundary between implanted and not-implanted \ce{SiO2} modifies the NC formation during the RTA step and that the ratio of circumference and area continuously tunes the mean NC size.

\section{Single NC-QD spectroscopy}
\begin{figure}[htb]
    \begin{center}
        \includegraphics[width=0.75\columnwidth]{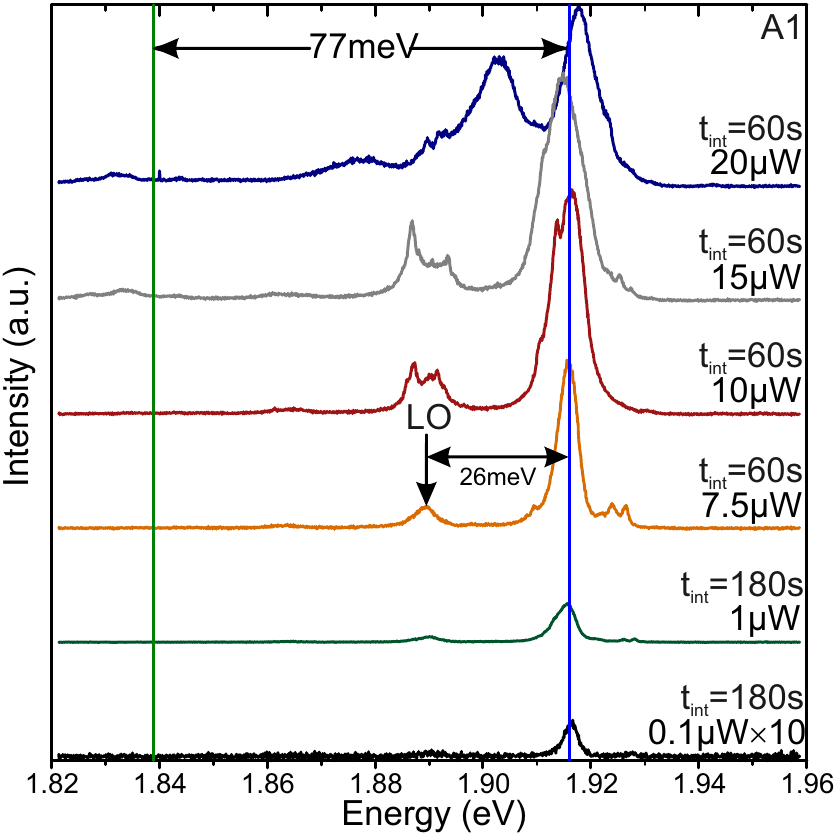}
        \caption{Single NC-QD emission - Laser-power dependent emission of a single NC-QD recorded from a $d=420\,{\rm nm}$ aperture.}
        \label{fig:4}
    \end{center}
\end{figure}
Finally we present first optical experiments preformed on single NC-QDs of the low fluence sample synthesized through $d=420\,{\rm nm}$ apertures (A1). In Fig. \ref{fig:4} we plot emission spectra recorded for different optical pump powers, increasing from $P_{\rm laser}= 100\,{\rm nW}$ (bottom spectrum) to $P_{\rm laser}= 20\,\mu{\rm W}$ (top spectrum). For low optical pump powers, the spectrum consists of a single emission line at $E_0=1.916\,{\rm eV}$. We attribute this emission line as arising from single exciton recombination, $\rm X=1e+1h$, of the NC-QD. The energy shift $E_0-E_{gap,bulk}=77\,{\rm meV}$ above the bulk energy gap of \ce{CdSe} suggests that this NC is in an intermediate confinement regime  and the measured confinement energy corresponds to a NC diameter of $d_{NC}= 11\pm1\,{\rm nm}$ \cite{Yu2003}. As the optical pump power increases the peak broadens significantly and its intensity increases. Furthermore, it develops additional side peaks at lower energies and a group of lines at $E\sim 1.927\,{\rm eV}$ appears. The group of lines at lower energies $E\sim 1.89\,{\rm eV}$ most likely originates from a second, optically active NC-QD within the implanted area. At the highest pump powers the signal is spectrally shifted first to lower and finally to higher energies indicating that at optical emission of these NC-QDs is sensitive to optical carrier generation rates.  This assumption is further supported by the relatively broad linewidth of ${\rm FWHM}\sim 4\,{\rm meV}$ at low pump powers. Due to the long integrations times ranging between $t_{int}= 60$-$180\,{\rm s}$ used to acquire the spectra in Fig. \ref{fig:4}, processes occurring on faster timescales cannot be resolved. \\

\begin{figure}[htb]
    \begin{center}
        \includegraphics[width=0.905\columnwidth]{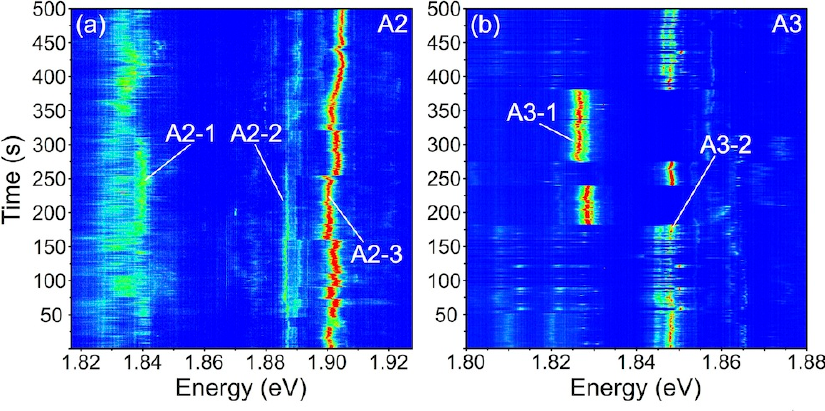}
        \caption{Spectral diffusion and intermittency of single NC-QDs - Time series of $t_{int}= 1\,{\rm s}$ spectra from aperture A2 (a), $d=820\,{\rm nm}$ and A3 (b), $d=630\,{\rm nm}$.}
        \label{fig:5}
    \end{center}
\end{figure}

We characterized the optical emission of individual NC-QDs by measuring its temporal evolution on a seconds time scale. For this experiment we selected two apertures (A2: $d=820\,{\rm nm}$, A3: $d=630\,{\rm nm}$) on the low fluence sample that exhibit particularly high PL intensities which allowed for $t_{int}= 1\,{\rm s}$ at $P_{\rm laser}= 20\,\mu{\rm W}$ for A2 and $P_{\rm laser}= 7\,\mu{\rm W}$ for A3. 
In Fig. \ref{fig:5} we compare two 500\,s time series from these apertures plotted in false-color representation. In these spectra we identify PL lines of different which we label A2/3-\textit{i}. For A2 these lines are centered at $E_{\rm A2-1}=1.837\,{\rm meV}$, $E_{\rm A2-2}=1.889\,{\rm meV}$ and $E_{\rm A2-3}=1.901\,{\rm meV}$ and at $E_{\rm A3-1}=1.827\,{\rm meV}$ and $E_{\rm A3-2}=1.847\,{\rm meV}$ for A3. In contrast to the spectra in Fig. \ref{fig:4} the line width of these lines is significantly reduced to ${\rm FWHM}\leq 0.7\,{\rm meV}$ and we expect further reduction for even lower optical pump powers as observed also for colloidal QDs at low temperatures \cite{Empedocles1996}. Furthermore, in both time series we observe both random spectral and intensity fluctuations of the isolated sharp emission lines. The amplitude of the spectral diffusion ranges between $\Delta E=35\,{\rm meV}$ for A3-2 and $\Delta E=15\,{\rm meV}$ for A2-1. This energy range points towards to this spectral diffusion being the origin of the large linewidths measured for long $t_{int}$. Such spectral diffusion and luminescence intermittency behaviors are well known for chemically synthesized colloidal QDs has been studied together with the underlying mechanisms in great detail since its first observation \cite{Nirmal1996,Empedocles1996,Neuhauser2000,Frantsuzov2008}. Most of these models are based on traps surrounding the QD which randomly capture and release photogenerated electrons. 
This difference of the amplitude of the spectral diffusion indeed points towards a pronounced impact of internal defects or defects at \ce{CdSe-SiO2} or the surrounding \ce{SiO2} matrix, which can dynamically charge and discharge. The local charge environment gives rise to small variations of the electronic levels of the NC-QD and the optical emission energies. The characteristic spectral and intensity blinking behaviors of each NC-QD are a direct consequence of the unique spatial distribution of defects. A closer examination of these statistical fluctuations of emission energy and intensity of A2 show that signals A2-2 and A2-3 and A3-1 and A3-2 are clearly correlated. Such correlated fluctuations are a clear fingerprint for emission signals originating from the same QD. In contrast there exist no such correlation between A2-1 and A2-2+3. These findings confirm that the correlated (uncorrelated) signals indeed stem from the same (different) NC-QD(s). This interpretation is further supported by the similar (different) amplitudes of the spectral diffusion.
A3-1 and A3-2 exhibit a digital switching which points to charging and discharging of an internal or interface defect. In contrast, A2-2 and A2-3 do not show such switching on the timescales accessible here, consistent with a fluctuating charge environment in the surrounding matrix which interacts weaker with the NC-QD.\\

\section{Summary and perspectives}
In summary, we demonstrated site-selective ion beam synthesis of \ce{CdSe} NCs in a \ce{SiO2} matrix using sub-micron implantation masks. The areal density, size and the resulting optical properties are controlled by the implantation fluence and the size of the apertures. For low fluences and small apertures we reach down to the regime of individual NC-QDs with confined energy states with discrete emission lines. The characteristics of the spectral diffusion and intermittency of single NC-QDs exhibit striking similarities to that of colloidal QDs at low temperatures\cite{Empedocles1996,Neuhauser2000}. Therefore, we conclude that our ion beam synthesized NC-QDs exhibit optical properties comparable to that of this established colloidal QDs. Their optical properties can be further improved adopting passivation strategies established for similar systems \cite{Min1996} to assess the statistics of the emitted photons and confirm single photon emission\cite{Michler2000}. Moreover, these optically active, bright emitters are ideally suited for coupled QD-nanocavity systems when hybridized with \ce{SiO2}-based photonic crystal structures\cite{Gong2010}. Our approach can be readily extended to other II-VI compound NC-QDs such as \ce{PbS}. This material is commonly used for near-infrared colloidal QDs emitting at $1.5\,\mu {\rm m}$ and is particularly tantalizing for CMOS integrated silicon photonics. Since ion beam implantation does not rely in planar substrates, our method can be applied to dope chemically inert, non-toxic \ce{SiO2} micro- and nanospheres with emitters for biomedical applications. The observed pronounced modification of the NCs' structural and optical properties by confining the implantation area to sub-micron dimensions is  equally appealing from an material science and solid-state chemistry point of view. It would be particularly interesting to transfer this approach to other types of NCs \cite{Meldrum2001,Pavesi2000,Zimmer2012} and investigate its applicability for different underlying solid-state chemical reactions.

\begin{acknowledgement}
We gratefully acknowledge financial support by the Deutsche Forschungsgemeinschaft (DFG) via the Cluster of Excellence {\it Nanosystems Initiative Munich} (NIM) and the Emmy Noether Program (KR3790/2-1). We thank Achim Wixforth for his continuous support of this project.
\end{acknowledgement}



\providecommand*\mcitethebibliography{\thebibliography}
\csname @ifundefined\endcsname{endmcitethebibliography}
  {\let\endmcitethebibliography\endthebibliography}{}

\begin{tocentry}
\includegraphics[width=0.85\columnwidth]{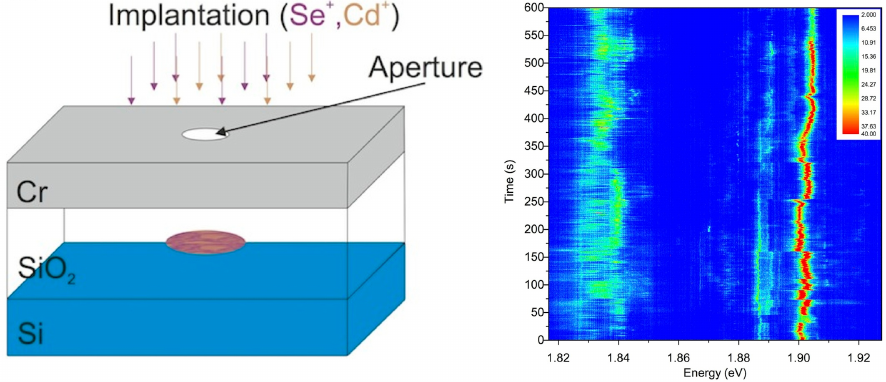}
\end{tocentry}

\end{document}